\documentclass[12pt]{article}
\usepackage{graphicx}
\usepackage{epsfig}
\usepackage{amssymb}

\input{epsf.sty}

\newcommand{\be}{\begin{equation}}
\newcommand{\ee}{\end{equation}}
\newcommand{\ba}{\begin{eqnarray}}
\newcommand{\ea}{\end{eqnarray}}
\newcommand{\bi}{\begin{itemize}}
\newcommand{\ei}{\end{itemize}}

\newcommand{\nn}{\nonumber}

\newcommand{\RR}{{\rm I\kern -.2em  R}} 
\newcommand{\eq}{Eq.~}
\newcommand{\eqs}{Eqs.~}
\newcommand{\fig}{Fig.~}

\def\lsi{\raise0.3ex\hbox{$<$\kern-0.75em\raise-1.1ex\hbox{$\sim$}}}
\def\gsi{\raise0.3ex\hbox{$>$\kern-0.75em\raise-1.1ex\hbox{$\sim$}}}
\newcommand{\lsim}{\mathop{\lsi}}

\def\none               {\multicolumn{2}{c|}{---}}

\begin{document}
 
\begin{titlepage}
\begin{flushright}
CERN-PH-TH/2006-081 \\
MS-TP-06-03
\end{flushright}
\begin{centering}
\vfill
 
{\bf\Large The chiral critical line of $N_f=2+1$ QCD \\ at 
                   zero and non-zero baryon density}

\vspace{0.8cm}
 
Philippe de Forcrand$^{1,2}$ and Owe Philipsen$^3$

\vspace{0.3cm}
{\em $^{\rm 1}$
Institut f\"ur Theoretische Physik,
ETH Z\"urich,
CH-8093 Z\"urich,
Switzerland\\}
{\em $^{\rm 2}$
CERN, Physics Department, TH Unit, CH-1211 Geneva 23,
Switzerland\\}
{\em $^{\rm 3}$
Institut f\"ur Theoretische Physik, Westf\"alische Wilhelms-Universit\"at M\"unster, Germany}

\vspace*{0.7cm}
 
\begin{abstract}
We present numerical results for the location of the chiral critical line
at finite temperature and zero and non-zero baryon density for QCD with $N_f=2+1$
flavours of staggered fermions on lattices with temporal extent $N_t=4$.
For degenerate quark masses, we compare our results obtained with the
exact RHMC algorithm with earlier, inexact R-algorithm results and find
a reduction of 25\% in the critical quark mass, for which the 
first order phase transition changes to a smooth crossover.
Extending our analysis to non-degenerate quark masses, we map out
the chiral critical line up to the neighbourhood of the physical
point, which we confirm to be in the crossover region.
Our data are consistent with a tricritical point at $(m_{u,d}=0,m_s\sim$500) MeV.

We also investigate the shift of the critical line with finite
baryon density, by simulating with an imaginary chemical potential
for which there is no sign problem.
We observe this shift to be very small or, conversely, the critical endpoint
$\mu^c(m_{u,d},m_s)$ to be extremely quark mass sensitive. 
Moreover, the sign of this shift is opposite to standard expectations. If confirmed on
a finer lattice, it implies the absence of a critical endpoint for physical QCD at small chemical potential, or another revision of the QCD phase diagram.
We critically examine earlier lattice determinations of the QCD critical point, and find them
to be in no contradiction with our conclusion.
Hence we argue that finer lattices are required to settle even the qualitative features of the 
QCD phase diagram.    

\end{abstract}
\end{centering}
 
\noindent
\vfill
\noindent
 

\vfill

\end{titlepage}
 

\section{Introduction}

Based on the property of asymptotic freedom, a fundamental prediction of QCD with three flavours of quarks is the transition from the familiar hadronic physics at low temperatures to a
regime of ``deconfined'' quark gluon plasma at high temperatures.
Whether this transition is characterised by singular behaviour of the partition function corresponding to a first or second order phase transition, or merely represents a smooth and analytic crossover between different dynamical regimes, depends crucially on the choice of the quark masses and the net baryon density specified by its chemical potential, $\mu_B$. In the following we shall assume the 
light quarks to be degenerate, $m_u=m_d=m_{u,d}$, and vary $m_s$ independently. 
We couple the quark chemical potential $\mu=\mu_B/3$ to the light quarks only, except for the degenerate case $N_f=3$, where all quarks 
are coupled.
The parameter space of the theory considered here is thus four-dimensional,  
$\{m_{u,d}, m_s, T, \mu_B\}$.

The first task in determining the phase diagram in this parameter space consists of finding
the (pseudo-)critical temperature $T_0(m_{u,d},m_s,\mu_B)$, defined e.g.~by the peak of
some susceptibility, which represents the boundary between the hadronic and plasma regimes. 
The independent variables $m_{u,d},m_s,\mu_B$ then span a 3d parameter space 
with regions of first order
phase transitions and analytic crossover separated by surfaces of second order phase 
transitions. In order to identify the order of phase transitions, and the location of the critical surfaces in particular, finite size scaling analyses are necessary.

\begin{figure}[tb]
\begin{center}
\vspace*{-3.0cm}            
{\rotatebox{0}{\scalebox{0.4}{\includegraphics{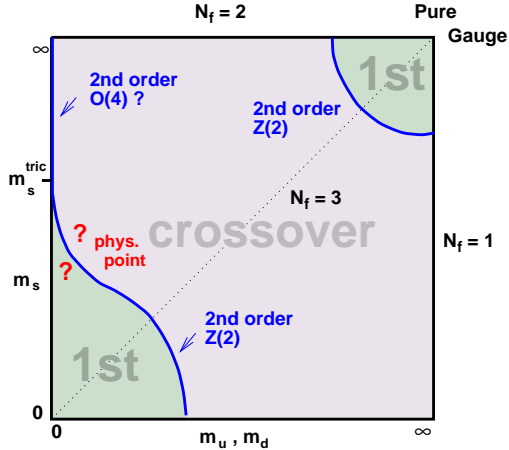}}}}
\end{center}
\caption[]{\em Schematic phase transition behaviour  of three flavour QCD for different choices of quark masses (from \cite{lp}), at zero density.}
\label{schem1}
\end{figure}

Let us first discuss the situation for $\mu_B=0$, shown schematically in \fig\ref{schem1} (for early references, see, e.g.~\cite{lp}).
Gauge invariant, local order parameters characterising the transition only exist in the extreme cases of zero or infinite quark masses, namely the chiral condensate and the Polyakov loop, respectively. These limiting theories thus must feature singular phase transitions, and one may write down effective theories of the Ginzburg-Landau type for the order parameters \cite{ptgen}.   
It is  numerically well-established that the phase transition is first order in the quenched \cite{nf3pt2}
limit, and there is strong numerical evidence for first order in the chiral  \cite{nf3pt1} limit.
Since first order phase transitions are robust against small variations of the parameters of the theory, 
the first order regions must extend by a finite amount into the quark mass plane. 
On the other hand, simulations have revealed smooth crossover behaviour for intermediate quark masses, which
implies second order boundary lines between the first order and crossover regions. 

In the case of heavy dynamical quarks, the relevant symmetry is the $Z_3$ center symmetry, and the weakening,
by the dynamical quarks, of the first-order transition which occurs in the Yang-Mills theory, is understood
qualitatively~\cite{hopping} and to a large extent quantitatively. Simulations have determined the second-order line to
correspond to a meson mass of about 2 GeV \cite{heavy_quarks}, and have confirmed the expectation that the universality class is that of the 3d Ising model.
In the case of light quarks, numerical simulations are more difficult, and 
very little is known quantitatively about the location of the second order boundary line.
The only point computed to some accuracy with standard staggered fermions is the chiral critical point 
\footnote{The superscript "$c$" here and in the following refers to "critical", not to charm.}
$m_{u,d}=m_s=m^c(\mu=0)\equiv m_0^c$ on the $N_f=3$ diagonal \cite{kls,clm,fp2}, which also was determined to belong to the 3d Ising universality class \cite{kls}.

While the statement about the universality class concerns infrared physics and thus is stable against cut-off effects, the location of the critical point in the physical mass plane   
turns out to be very strongly affected. To date calculations have been performed on lattices with 4 time-slices only ($N_t=4$, implying a lattice
spacing $a = \frac{1}{N_t T} \sim 0.3$ fm), but simulations with improved actions indicate values for $m^c_0$, and the associated pion mass,  which are  
considerably smaller than the standard action result \cite{kls}. 
Moreover, all these simulations used the so-called R-algorithm \cite{ralg}, which has stepsize errors
and therefore gives only approximate results in the absence of a careful extrapolation to zero stepsize.
In any case,
all current results are consistent with the physical point being in the crossover regime.

\begin{figure}[tb]
\vspace*{-3.5cm}  
\hspace*{-1cm}          
{\rotatebox{0}{\scalebox{0.7}{\includegraphics{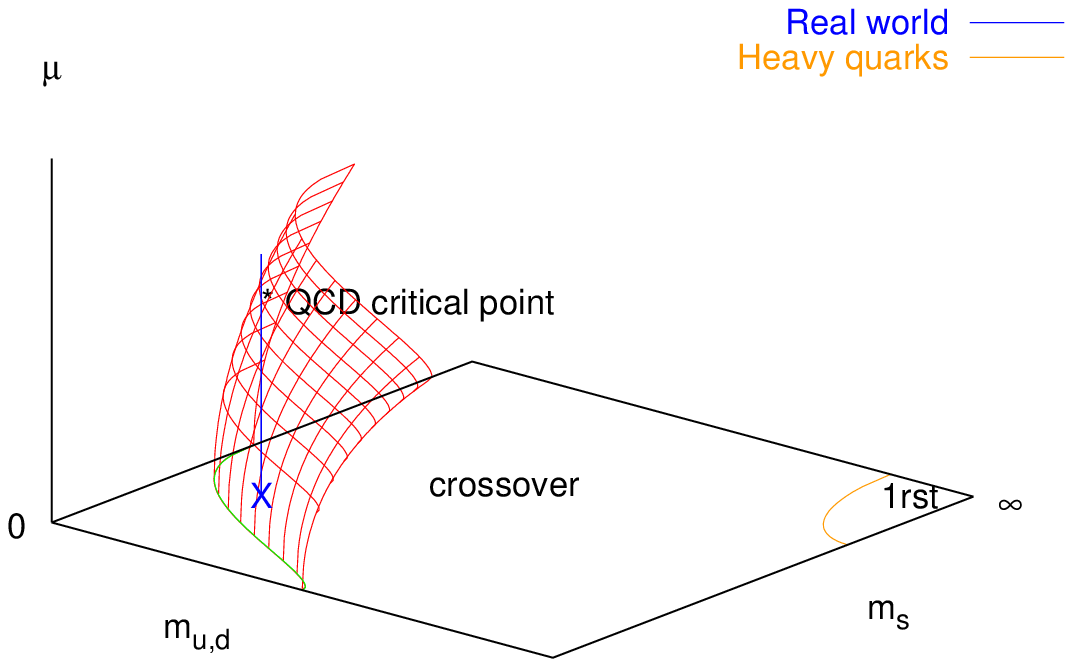}}}}
{\rotatebox{0}{\scalebox{0.7}{\includegraphics{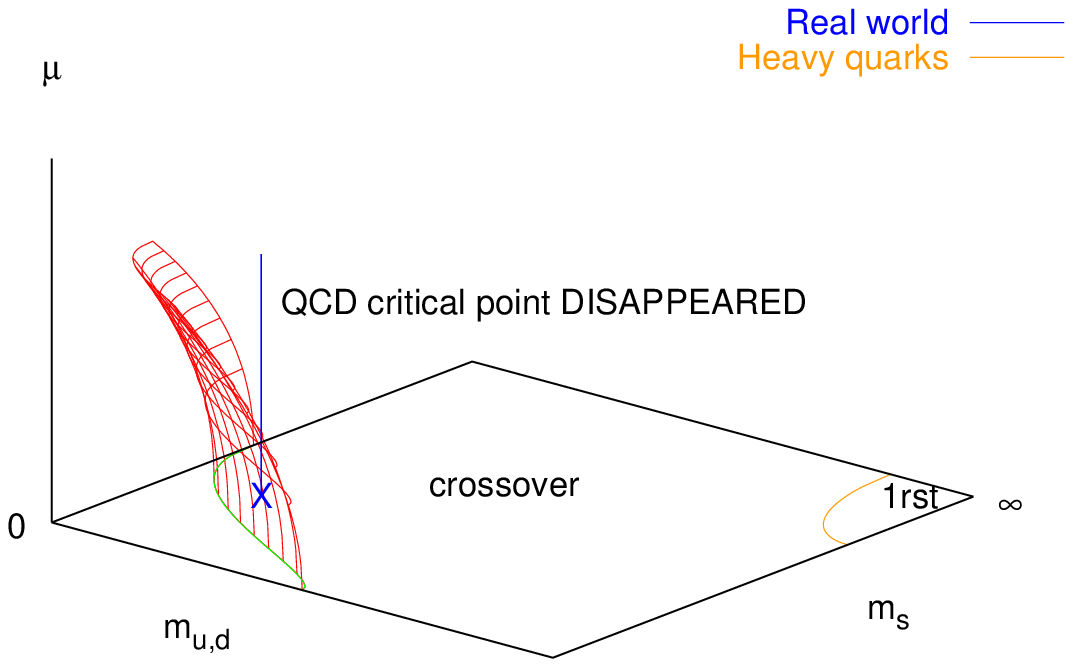}}}}
\vspace*{-1cm}
\caption[]{\em The chiral critical surface in the case of positive and negative curvature. If the physical point 
is in the crossover region for $\mu=0$, a finite $\mu$ phase transition will only arise in the scenario
with positive curvature. }
\label{schem}
\end{figure}

In the presence of a chemical potential the second order boundary lines turn into surfaces,
as indicated in \fig \ref{schem}. The qualitative features of the $(T-\mu)$ phase diagram now depend
crucially on the curvature at $\mu=0$, $d^2m^c/d\mu^2(0)$. The common expectation is that this
curvature is positive. 
Hence the physical point, once the chemical potential is increased, 
will be closer to the critical line, and intersect it for
a critical chemical potential $\mu^c$. For values larger than $\mu^c$ a first order phase transition is expected.
Clearly, this is not the case for negative curvature of the critical surface.

In this work we present a comprehensive numerical study mapping out the chiral critical line in simulations of  the standard staggered action on several lattices with $N_t=4$. Upon repeating the computation for the $N_f=3$ chiral critical point with the rational hybrid Monte Carlo (RHMC) algorithm \cite{rhmc}, which is free of finite step size errors, we
find that the bare quark mass $am^c_0$ is reduced by 25\%, and the physical pion mass by 10\%, compared to the 
accepted values determined previously using
the R-algorithm. We then extend our simulations to cover a wide range of quark masses, mapping out the critical line up to the neighbourhood of the physical point. In agreement with expectations, the physical point is found to be on the crossover side of the boundary. 
Assuming O(4) behaviour for the $N_f=2$ chiral limit, the fit to our critical line can be extrapolated to the $m_{u,d}=0$ axis consistently with the required $O(4)$ scaling behaviour, putting the tri-critical point in that scenario 
(see \fig \ref{schem1}) around $m_s^{tric}/T\sim 2.8$. However, non-O(4) behaviour is not excluded by our data.
Our results should also provide a testing ground and input for analytic attempts to determine the
critical line from effective theories based on universality arguments \cite{eff} 
(for a review, see \cite{effrev}).

In a second set of simulations, we repeat the analysis for an imaginary baryon chemical potential
$\mu_B/(iT) = 2.4$ and determine the corresponding shift of the critical line, following the strategy already used in \cite{fp2}.
Together with additional imaginary $\mu$ simulations for the $N_f=3$ case,
this allows for a determination of the curvature of the critical surface at $\mu_B=0$, which can be readily continued to real values of $\mu_B$. We find this curvature to be
{\em negative}, as illustrated in \fig \ref{schem} (right).  
In the $(T-\mu)$ phase diagram this implies that the critical endpoint moves to smaller
$\mu$ with growing quark mass, until it disappears entirely for physical quark masses. 
This is contrary to customary expectations, and in contradiction with the results of \cite{fk2}, obtained at the same lattice spacing and with the same action, but using the R-algorithm and a different numerical approach. 
Clearly, a careful study of systematic errors, due in particular to the very coarse lattice spacing, is needed.
Still, if the physical point of QCD is indeed in the crossover region at $\mu_B=0$, our finding would imply that the transition will remain an analytic crossover
also for any finite $\mu_B\lsim 500$ MeV,
placing a possible QCD critical point at much larger values of $\mu_B$. 
Preliminary results to this extent have already been given in
 \cite{op}.

After summarising the properties of QCD at imaginary $\mu$ and introducing the Binder cumulant as our observable for the order of the phase transition in 
Secs.~\ref{sec:imag}, \ref{sec:binder}, respectively, we begin our analysis in Sec.~\ref{sec:step} with a thorough discussion of step size effects for $N_f=3$ and a comparison of results from the R- and RHMC-algorithms. 
The computation of the chiral critical line for $\mu_B=0$ and $\mu_B/(iT)=2.4$ is presented in 
Sec.~\ref{sec:line}, which also discusses the resulting new scenario for the ($T-\mu_B$) phase diagram of physical QCD. An assessment of systematic uncertainties is contained in 
Sec.~\ref{sec:sys}, along with our conclusions.

\section{QCD at imaginary $\mu$ \label{sec:imag}}

\begin{figure}[t]
\begin{center}
\includegraphics*[width=0.45\textwidth]{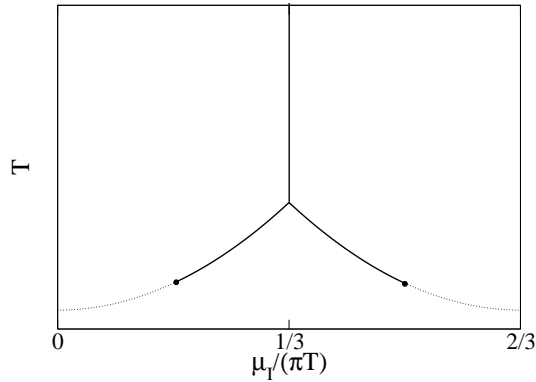}
\caption[]{\em Schematic phase diagram for QCD at imaginary chemical potential. The diagram is periodically repeated for larger values of $\mu_i$.
The heavy lines indicate first order transitions, the thin lines crossovers.
The nature of the temperature-driven transition depends on the parameters
or the theory ($N_f$, quark masses).}
\label{ischem}
\end{center}
\end{figure}

In order to study the phase diagram \fig\ref{schem} at finite baryon density, we employ simulations
at imaginary chemical potential $\mu=i\mu_i$, where the fermion determinant is positive, followed by analytic continuation, as discussed in detail in previous work \cite{fp1,fp2}. 
To render the paper self-contained, we briefly recall some points needed in the sequel. 
The QCD partition function at finite baryon chemical potential $\mu_B=3\mu$ is even under reflection 
$\mu\rightarrow -\mu$. Moreover, it is periodic in the imaginary direction, with period $2\pi/N_c$ for $N_c$ colours \cite{rw}, i.e.~$Z(\mu_r/T,\mu_i/T)=Z(\mu_r/T,\mu_i/T+2\pi/3)$.
Because of the fermionic boundary conditions, this symmetry implies that a shift in $\mu_i$ by $2\pi/3$ is exactly compensated by a $Z(3)$-transformation, so that $Z(3)$ transitions take place between neighbouring centre sectors for all $(\mu_i/T)_c=\frac{2\pi}{3} \left(n+\frac{1}{2}\right), n=0,\pm1,\pm2,...$. It has been numerically verified that these transitions
are first order for high temperatures and a smooth crossover for low temperatures \cite{fp1,el1}, as in \fig \ref{ischem}. Hence, the first of these transitions limits the radius of convergence for analytic continuation to the first sector for most observables. With a pseudo-critical temperature of $T_0\sim 170$ MeV, our accessible physics range is thus $\mu_B\lsim 500$ MeV.

Within this first sector, observables can be simulated at imaginary $\mu=i\mu_i$. The results
may be fitted by truncated Taylor series 
$ \langle O \rangle = \sum_n^N c_n(\mu_i/T)^{2n}$, whose 
convergence can be tested by inspection. Analytic continuation of successful fits is then trivial.

A remarkable finding of previous work inspecting Taylor series is that, within $\mu_B\lsim 500$ MeV, convergence is rapid
and screening masses \cite{hlp,taro}, the pressure \cite{swabi} as well as the pseudo-critical temperature \cite{fp1,fp2} are all well described by the leading term $\sim \mu^2$. 
This becomes plausible by noting that at finite $T$ the natural expansion parameter is $\mu/(\pi T)$ rather than $\mu/T$ \cite{hlp,taro}.
Hence we write the pseudo-critical temperature, separating the hadronic from the plasma phase, and the   
critical quark mass marking the boundary between first order and crossover behaviour, as
\ba
\frac{T_0(m_{u,d},m_s,\mu)}{T_0(m_{u,d},m_s,\mu=0)}&=&
1 +  b_1(m_{u,d},m_s) \left(\frac{\mu}{\pi T}\right)^2+\ldots \\
\frac{m_{u,d}^c(m_{s},\mu)}{m_{u,d}^c(m_{s},\mu=0)}&=&
1 +  c_1(m_{s}) \left(\frac{\mu}{\pi T}\right)^2+\ldots
\label{c1}
\ea
The choice to treat $m_{u,d}$ as the independent variable parametrizing the critical line in \eq (\ref{c1}) reflects our practical procedure, namely to fix $m_s$ and then scan in $m_{u,d}$, 
because the critical line is a steeper function of the latter.
We shall determine the coefficients $b_1,c_1$ quantitatively and free of step size errors for the $N_f=3$ theory, and provide the sign of $c_1$ along the whole chiral critical line for $N_f=2+1$.

\section{Universality and the Binder cumulant \label{sec:binder}}

There are different ways to investigate and exhibit critical behaviour of the theory along the critical line, such as finite size scaling (FSS) of susceptibilities,  of Lee-Yang zeroes or of the Binder cumulant.
The Binder cumulant \cite{binder} offers various advantages for our particular study. It is defined as
\be
B_4(m^c,\mu^c)=\frac{\langle(\delta X)^4\rangle}
{\langle(\delta X)^2\rangle^2},
\ee
with the fluctuation $\delta X= X-\langle X\rangle$ of some observable $X$ around its mean value,
evaluated at the pseudocritical coupling (determined by a peak in $\langle(\delta X)^2\rangle$ or a zero in
$\langle(\delta X)^3\rangle$). In practice we study $X_i=\bar{\psi_i}\psi_i, i=1,2$ for the two quark masses $m_{u,d}, m_s$.
In the infinite volume limit the Binder cumulant behaves discontinuously, assuming the values 1 in a first order regime,
3 in a crossover regime and some critical value reflecting the universality class at a second order critical point. 
On a finite volume the discontinuities are smeared out and flattened,  so that $B_4$ passes continuously through the critical value.  The location of the critical point in parameter space will then be displaced by some finite volume correction. 

In \cite{kls} the Binder cumulant was chosen as observable because its critical value for the expected  Z(2) universality  class is distinct from those corresponding to other symmetries like O(2), O(4) etc.
In this way Z(2) scaling for the $N_f=3$ chiral critical point was clearly established in \cite{kls}.
Once the universality class is ascertained, the Binder cumulant allows to approximately map out the 
critical line on a fixed lattice size, by scanning the parameter space for the line on which $B_4$ is 
held constant at its critical value.
In practice this is best achieved by holding one quark mass fixed, and scanning in the other.
In the small $m_{u,d}$ regime the critical line turns out to be a steep 
function $m_s^c(m_{u,d})$,
and thus we choose to scan in the light quark mass while keeping $m_s$ fixed. 
In the neighbourhood of a critical point $B_4$ can then be expanded in a Taylor series,
\be
B_4(am_{u,d},am_s,a\mu)=\sum_{n,m}b_{nm}(m_s) \left(am_{u,d}-am_{u,d}^c(m_s)\right)^n
(a\mu)^m,
\label{bseries}
\ee
with $b_{00}(m_s)\rightarrow 1.604$ for $V\rightarrow \infty$.
For $N_f=3$ there is only one mass variable $m_{u,d}=m_s=m$ in the above expression, and the mass dependence of the coefficients disappears. 

For large volumes the approach to the thermodynamic
limit is governed by universality.
Near a critical point the correlation length diverges
as $\xi\sim r^{-\nu}$, where $r$ is the distance to the critical point
in the plane of temperature and magnetic field-like variables, and $\nu\approx 0.63$
for the 3d Ising universality class. In practice, we first find the pseudo-critical gauge coupling
$\beta_0$ for a given pair $(m_{u,d},m_s)$,
and then compute $B_4$ for those parameter values.
Since $\beta$ is tuned to $\beta_0$ always, we have $r=|am_{u,d}-am_{u,d}^c(m_s)|$.
$B_4$ is a function of the dimensionless ratio $L/\xi$, or equivalently
$(L/\xi)^{1/\nu}$. Hence  one expects
the universal scaling behavior
\be \label{scale}
B_4\left((L/\xi)^{1/\nu}\right)=B_4\left(L^{1/\nu}(am_{u,d}-am_{u,d}^c(m_s)\right)\;.
\ee

\section{$N_f=3$ without step size errors \label{sec:step}}

The algorithm most widely used in simulations of finite temperature QCD in the staggered formulation, both standard and improved,  is the R-algorithm \cite{ralg}, which was also employed in previous studies of the chiral critical point for $N_f=3$ \cite{kls,clm,fp2}. 
As pointed out in \cite{ralg}, the ``correct'' usage of the R-algorithm consists of performing simulations for various choices of decreasing stepsizes, followed by an extrapolation to zero stepsize.
However, in practice usually a shortcut avoiding the extrapolation is applied: for some reference value of the quark masses, choose a step size for which the step size error is smaller than the typical  statistical error of the simulation. The molecular dynamics of the R-algorithm then suggests
to keep the ratio of quark mass $a m_q$ and step size $\delta \tau$ constant, i.e.~adjust the stepsize accordingly when the quark mass is reduced.
A typical choice is $\delta \tau = \frac{1}{2} a m_q$, 
although in many cases $\frac{2}{3} a m_q$ or even $a m_q$ have been adopted.

While this procedure has been followed successfully in the intermediate quark mass regime, it breaks down for small quark masses, where the linear relation no longer appears to hold and the step size needs to be decreased faster than proportionally. Furthermore, in a study of the QCD phase transition at finite isospin chemical potential it has recently been demonstrated that a finite step size leads to a systematic underestimate of $B_4$ \cite{ks}. Hence too coarse a stepsize can fake a first order transition, when the zero step size result really represents a crossover behaviour.  

\subsection{The order of the transition: R- vs.~RHMC algorithm}

In order to control this important source of systematic error, we have returned to our investigation of the $N_f=3$  critical point at $\mu_B=0$~\cite{fp2}, this time with the RHMC-algorithm. 
This algorithm has no stepsize errors and is exact.
For a discussion of the algorithm and numerical test results, see \cite{r_vs_rhmc}.
Our numerical procedure to compute the Binder cumulant is as follows.
For each set of fixed quark mass and chemical potential, we determine $\beta_0$ by interpolating from a range of typically 3-4 simulated $\beta$-values by Ferrenberg-Swendsen reweighting \cite{fes}.
For each simulation point 50k-200k RHMC trajectories have been accumulated, 
measuring the gauge action, the Polyakov loop and up to four powers 
of the chiral condensate after each trajectory. Thus, the estimate of 
$B_4$ for one mass value consists of at least 200k, and the estimate of
the critical mass of at least 800k trajectories.
\begin{figure}[t]
\hspace*{-1cm}
\includegraphics*[width=0.55\textwidth]{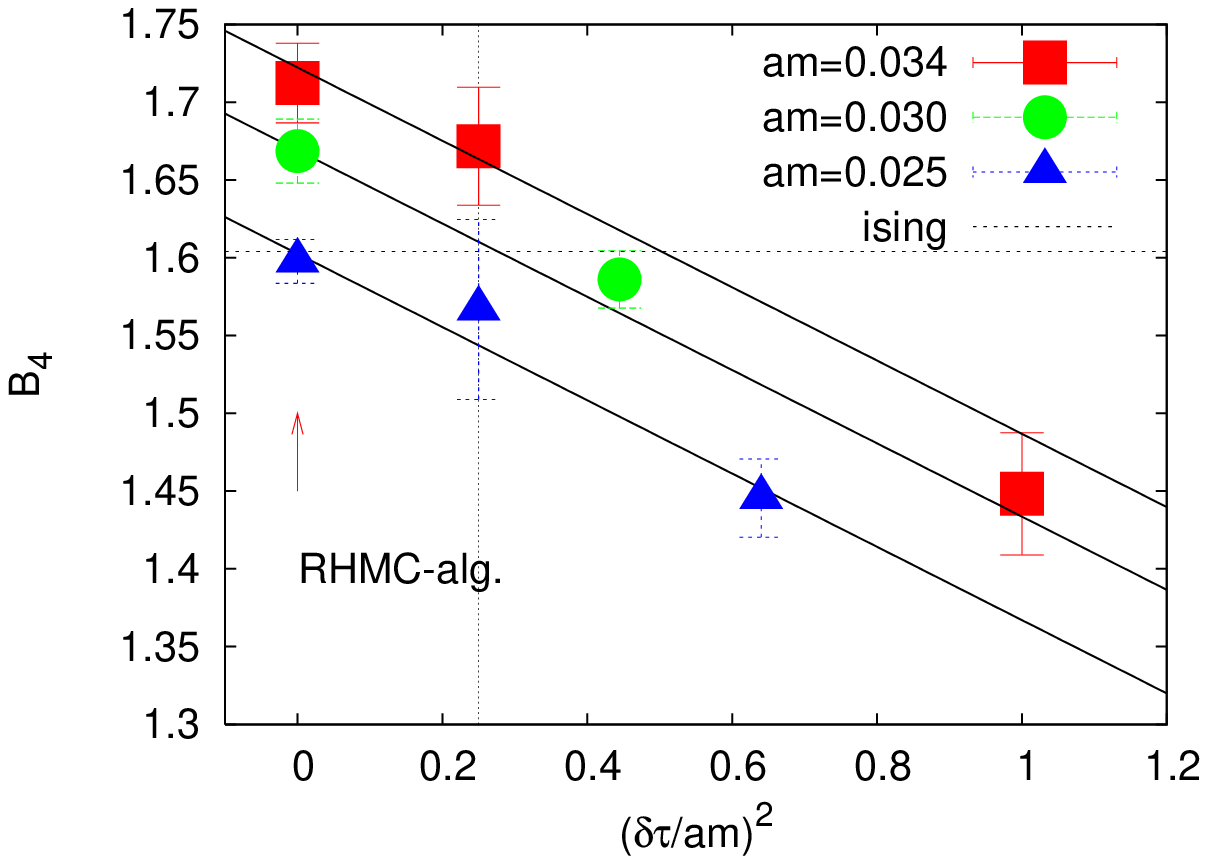}
\includegraphics*[width=0.55\textwidth]{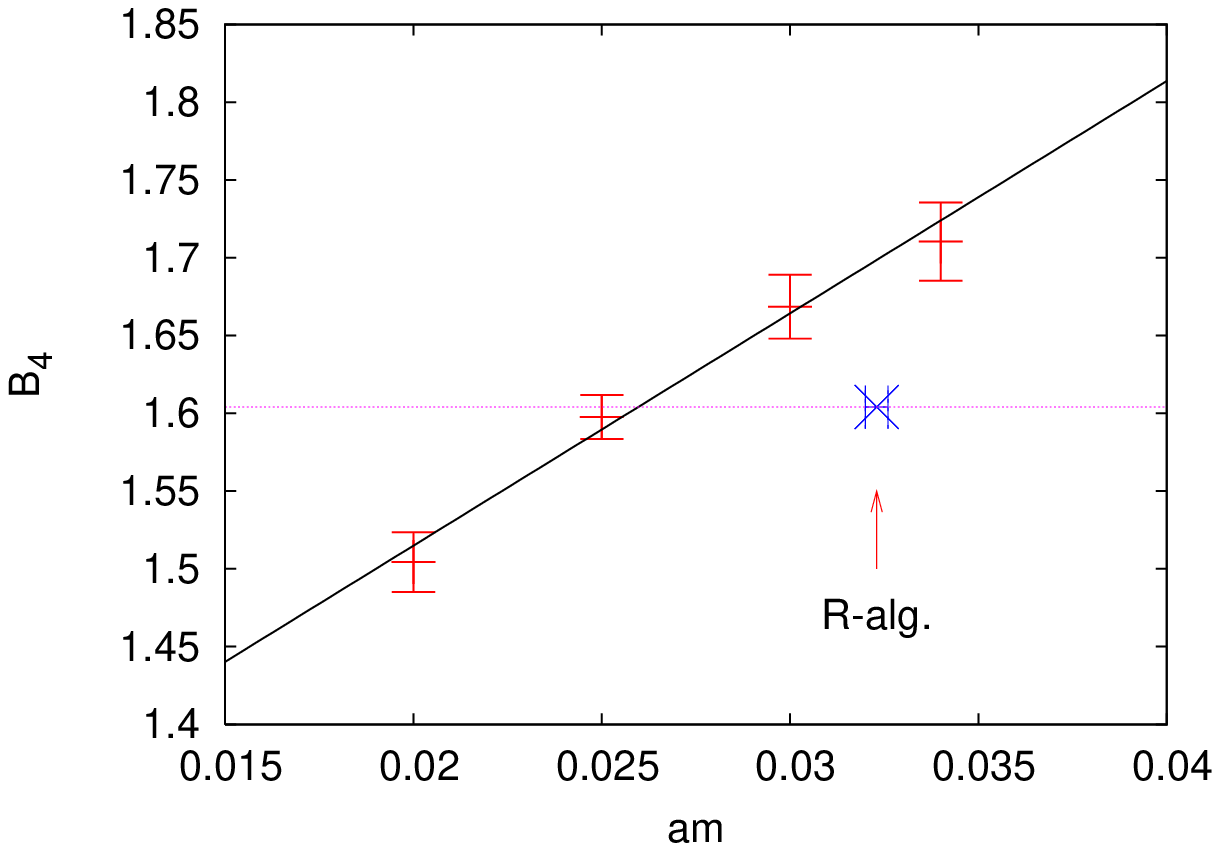}
\caption[]{\em Left: Comparison of the Binder cumulant computed with the 
RHMC algorithm (leftmost data) and the zero stepsize extrapolation of the R-algorithm. The solid lines
represent a common fit to all data, 
the vertical line marks the commonly used R-algorithm step size when no extrapolation is performed.
Right: Determination of $m^c(\mu=0)=m^c_0$ with the RHMC algorithm. The arrow marks the result from the R-algorithm~\cite{fp2}.}
\label{oldnew}
\end{figure}

\fig\ref{oldnew} (left) shows results of this first study, comparing measurements of $B_4$ from the RHMC algorithm with those obtained from the R-algorithm at various step sizes. The figure confirms the finding from \cite{ks} that decreasing the step size leads to an increase in the values of the Binder cumulant. It also constitutes a useful test of the RHMC algorithm, whose results indeed correspond to the zero step size limit of the R-algorithm. We note that for our smallest quark masses studied, $am=0.005$, the RHMC algorithm runs over 20 times faster than the R-algorithm at the commonly
applied step size. Since the latter also requires runs at several step sizes for the extrapolation, the RHMC is thus considerably more economical in producing results free of step size errors.

\subsection{The $N_f=3$ critical quark mass at $\mu=0$}

In order to eliminate step size errors,  
we now proceed to repeat the calculation of the critical quark mass $m^c_0$ in the three-flavour theory by  means of the Binder cumulant, this time with the RHMC algorithm.
The result obtained on an $8^3$ lattice is shown in \fig\ref{oldnew} (right). Qualitatively the behaviour is the same as previously,
with $B_4$ growing from first order behaviour through its critical Ising value to crossover with increasing quark mass, which can be fitted to leading order in the quark mass.
However, the critical Ising value is now obtained at a bare mass of $am_0^c=0.0260(5)$,
which is about 25\% smaller than the value $am_0^c\approx 0.033$ quoted by all previous work using the R-algorithm \cite{kls,clm,fp2}. 

One may ask whether this change affects bare quantities only, while the R and RHMC algorithms
probe the same physics. To study this issue, we measured the zero-temperature hadron spectrum
at the parameters $(\beta_c,am_c)_{RHMC}$ using RHMC, and compared with the same exercise performed in \cite{kls} at the parameters $(\beta_c,am_c)_R$ using the R-algorithm.
For the pion, which is the most accurately determined, the ratio $m_\pi/T_0$ changes from
1.853(1) (R \cite{kls}) to 1.680(3) (RHMC). 
This reduction of 10\% in the pion mass corresponds to a change of 20\%
in the renormalized quark mass, very near the observed 25\% change in the bare quark mass.
Therefore, replacing the R by the RHMC algorithm corrects a large error in the {\em physical}
values of the critical parameters. The correction should be even larger for
smaller $m_{u,d}$ masses.
We conclude that for the study of the QCD phase transition in the region of physical 
quark masses, step size errors in the Monte Carlo algorithm can lead to a qualitatively different picture
at fixed parameter values, and the use of an exact simulation algorithm is mandatory.

\begin{figure}[t]
\begin{center}
\includegraphics*[width=0.6\textwidth]{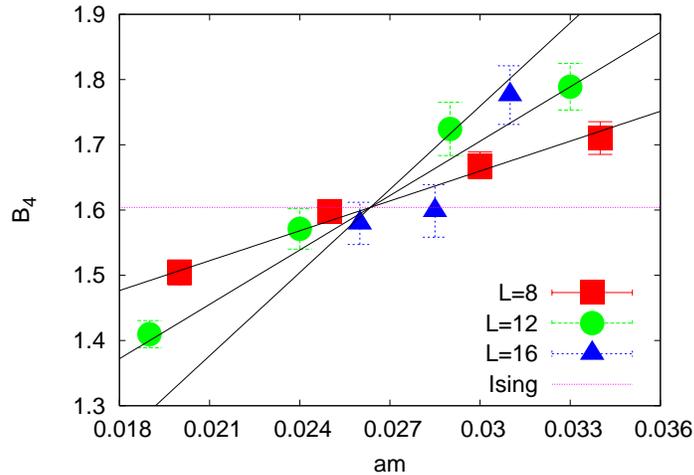}
\caption[]{\em Finite size scaling of the Binder cumulant for $N_f=3$. The lines represent a common fit 
to the data, according to \eq (\ref{fssfit}).}
\label{fss}
\end{center}
\end{figure}
Our results so far have been obtained on a single spatial volume $8^3$. The next task is to study  the FSS behaviour and uncover possible finite volume effects. This is particularly important since large finite size corrections were reported in a recent investigation of the chiral critical point at finite density in the Taylor expansion of susceptibilities \cite{gg}.
\fig\ref{fss} shows data obtained for three lattice sizes with $L=8,12,16$. 
According to \eq (\ref{scale}) and the corresponding discussion, in the scaling region near a critical point $B_4$ should be described by 
\be
B_4(m,L)=b_0+b L^{1/\nu}(m-m^c_0),
\label{fssfit}
\ee
with $b_0=1.604$ and $\nu=0.63$ for 3d Ising universality.
We have checked for finite volume effects by fixing $b_0$ to the Ising value and fitting for $b,\nu$ and $m^c_0$. With a $\chi^2$ of 0.74 per d.o.f., we obtain $am_0^c=0.0263(3)$ 
consistent with our result from $L=8$ only, and 
$\nu=0.67(13)$, which is consistent with the Ising exponent.
We conclude that for $N_f=3$ the Binder cumulant is close to thermodynamic scaling for lattice sizes $L\geq 8$, and hence finite volume effects are under control in this calculation.

\subsection{The pseudo-critical temperature for $N_f=3$ at finite $\mu$}

\begin{figure}[t]
\begin{center}
\includegraphics*[width=0.6\textwidth]{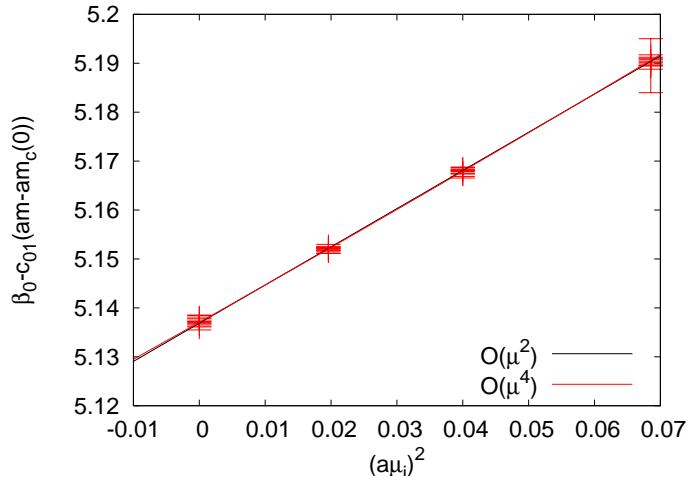}
\caption[]{\em The pseudo-critical coupling $\beta_0$ for $N_f=3$ on an $8^3\times 4$ 
lattice as a function of quark mass and imaginary chemical potential. 
For each value of $\mu_i$, $\beta_0$  is determined from 4 data points 
corresponding to masses $am=0.02-0.034$. The two lines represent the fits from Table \ref{fits}.}
\label{betac}
\end{center}
\end{figure}

On the lattice,  $T_0$ is determined from the pseudo-critical gauge coupling, which we define 
as the location of the peak of the plaquette susceptibility.
On any finite volume it can be expanded as a double series in mass and chemical potential around the three flavour critical point $m^c_0$,
\be
\beta_0(a\mu,am)=\sum_{k,l=0} c_{kl}\, (a\mu)^{2k}\, (am-am^c_0)^l.
\label{beta}
\ee
\fig \ref{betac} shows the measured values of $\beta_0(am,a\mu_i)$ for four different imaginary 
chemical potentials spanning the whole $\mu_i$ range up to the first $Z(3)$-transition. For each value of $a\mu_i$,
four quark masses in the range $0.02 \leq am \leq 0.34$ have been simulated.
\begin{table}[t]
\begin{center}
\hspace*{-1cm}
\begin{tabular}{|*{5}{r@{.}l|}l|}
\hline
\multicolumn{2}{|c|}{$c_{00}=\beta_0(0,m^c_0)$} &
\multicolumn{2}{c|}{$c_{10}, (\mu^2)$} &
\multicolumn{2}{c|}{$c_{20}, (\mu^4)$} &
\multicolumn{2}{c|}{$c_{01}, (m)$} &
\multicolumn{2}{c|}{$\chi^2/{\rm dof}$} \\
\hline
 5&1369(3) & 0&781(7)  & \none    & 1&94(3)   & 1&28 \\
 5&1371(3) & 0&759(22) & 0&33(32) & 1&95(4)   & 1&27 \\
\hline
\end{tabular}
\caption{ \label{fits} 
  {\em 
Fits of the Taylor expansion $\beta_0(am,a\mu)$,
\eq (\ref{beta}), to our data in \fig\ref{betac}.}}
\end{center}
\end{table}
As in our previous study, we obtain good fits retaining the leading terms only, as shown in Table \ref{fits}.
In particular, the term quadratic in the chemical potential is now sufficient to describe the data all the way out
to $\mu/T=1$, the quartic coefficient being consistent with zero.
Using the two-loop beta function,
this translates to a pseudo-critical temperature of
\be
\frac{T_0(m,\mu)}{T_0(m^c_0,0)}= 1 
+ 2.111(17) \left(\frac{m-m^c_0}{\pi T_0}\right) 
-0.667(6)\left(\frac{\mu}{\pi T_0}\right)^2
+0.23(9)\left(\frac{\mu}{\pi T_0}\right)^4+\ldots
\label{nf3temp}
\ee
Again, we note a shift of up to 10\% in the coefficients compared to the R algorithm results \cite{fp2}.
\eq (\ref{nf3temp}) has to be considered with some caution, since it is well known that the two-loop beta function is rather inaccurate at our coarse lattice spacing. 
The effect of the non-perturbative beta function is to increase the absolute
values of the coefficients $A, B$, perhaps by up to a factor of 2 \cite{betafunc}.

\subsection{Quark mass dependence of the critical point for $N_f=3$ \label{sec:c1}}

In order to detemine the order of the transition, we now repeat the previous procedure to find
how the critical bare quark mass $m^c(\mu)$ changes with imaginary chemical potential. 
As in the case of the pseudo-critical temperature, we express this by a Taylor series \eq (\ref{c1}):
\be
\frac{m^c(\mu)}{m^c(\mu=0)}=1 +  c_1 \left(\frac{\mu}{\pi T}\right)^2+\ldots,
\label{c03}
\ee
in order to be able to continue to real $\mu$.
Inversion
of this function will then give the location of the critical point as function of the quark mass, $\mu^c(m)$.
In practice, at a finite lattice spacing we are dealing with the expansion in lattice units,
\be
am^c(a\mu) = am^c(a\mu=0)+c_1'(a\mu)^2+c_2' (a\mu)^4 +\ldots
\label{c03lat}
\ee
A crucial point is that, for fixed temporal lattice extent $N_t$, the lattice spacing entering the dimensionless 
$am^c(\mu)$ and $am^c(0)$ is different, since $T_0(m_c(\mu),\mu)=1/(N_t a(\mu))$ 
depends on $\mu$. The relation of the 
leading coefficient to its continuum counterpart is thus given by
\be
c_1=\frac{1}{m^c_0}\,\frac{dm^c}{d(\mu/\pi T)^2}=\frac{\pi^2 }{N_t^2}\frac{c_1'}{a m^c_0}
+\frac{1}{T_0(m^c_0,0)}\frac{dT_0(m^c(\mu),\mu)}{d(\mu/\pi T)^2},
\ee  
or in terms of $A=2.111(17),B=-0.667(6)$ from \eq (\ref{nf3temp}) and $c'_1$ from \eq (\ref{c03lat}),
\be
c_1=\left(\frac{\pi^2}{N_t^2}\frac{c'_1}{a m^c_0}+B\right)\left(1-A\frac{m^c_0}{\pi T}\right)^{-1}.
\label{conv}
\ee

The coefficients $c'_i$
are extracted from our data for $B_4$ obtained at imaginary $\mu=i\mu_i$, by fitting to a double expansion about the known critical point at $m_c(\mu=0)$,
\be \label{Bexp}
B_4(am,a\mu)=\sum_{n,l} b_{nl}\, (am-am^c_0)^n(a\mu)^{2l}\;.
\ee
The leading coefficients $c'_i$ are then obtained as
\ba 
c'_1=\frac{d\,am^c}{d(a\mu)^2}&=&-\frac{\partial B_4}{\partial (a\mu)^2}
\left(\frac{\partial B_4}{\partial am}\right)^{-1}=-\frac{b_{01}}{b_{10}}\;,\label{der1}\\
c'_2=\frac{d^2\,am^c}{d[(a\mu)^2]^2}&=&-\frac{\partial^2B_4}{\partial(a\mu)^2}
\left(\frac{\partial B_4}{\partial am}\right)^{-1}
+\frac{\partial B_4}{\partial (a\mu)^2}\left(\frac{\partial B_4}{\partial am}\right)^{-2}
\frac{\partial^2 B_4}{\partial (a\mu)^2\partial am}\nn\\
&=&-\frac{b_{02}}{b_{10}}+\frac{b_{01} b_{11}}{b_{10}^2}
\label{der2}
\ea
For the actual analysis 
it is thus convenient to reparametrise the second order expansion of $B_4$ as
\ba 
B_4(am,a\mu)&=&1.604+b_{10}\left[am-am^c_0-c_1'(a\mu)^2\right]+b_{20}(am-am^c_0)^2\nn\\
&&-b_{10}\left[(c_2'-c_1'C)(a\mu)^4+C(am-am^c_0)(a\mu)^2\right], 
\label{b4f}
\ea
with $C=-b_{11}/b_{10}$, and fit the data via the parameters $m^c_0,b_{10},b_{20},c_1',c_2',C$.

\begin{figure}[t]
\hspace*{-1cm}
\includegraphics*[width=0.55\textwidth]{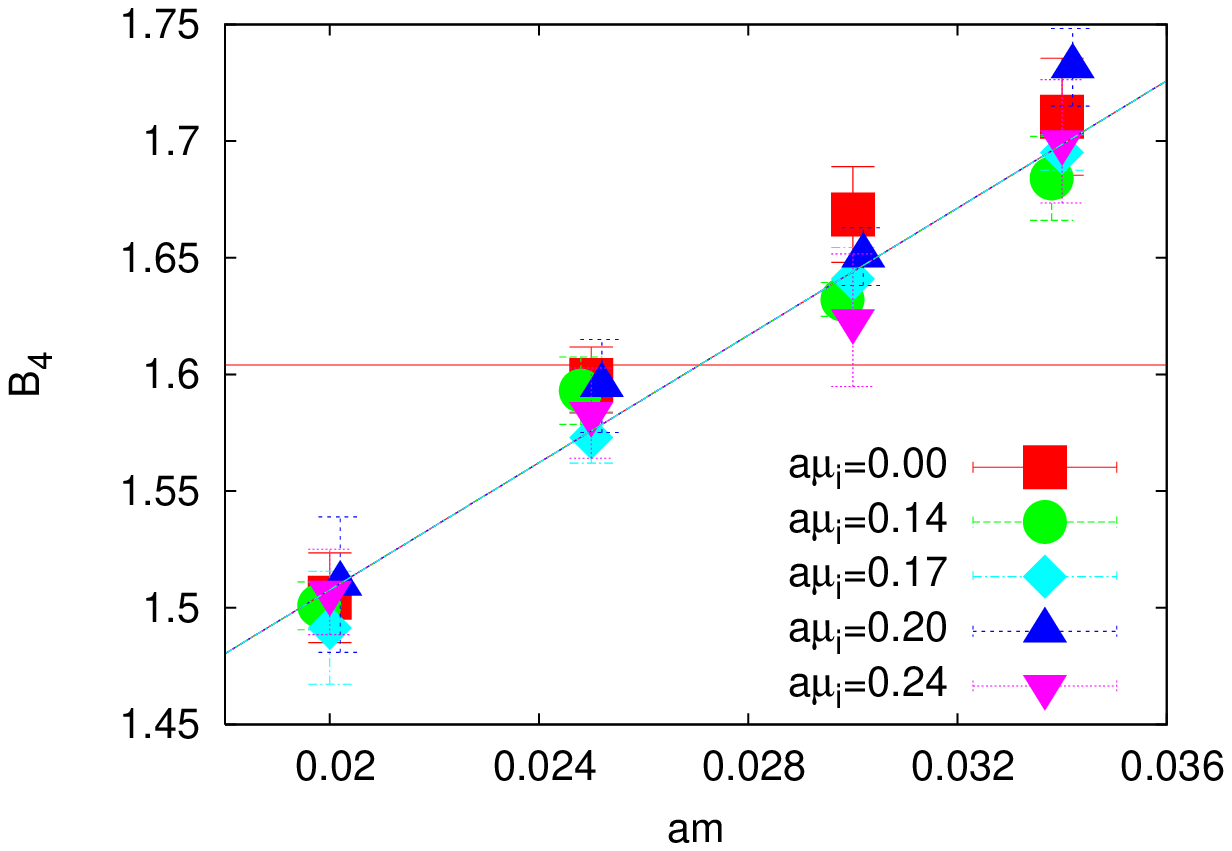}
\includegraphics*[width=0.55\textwidth]{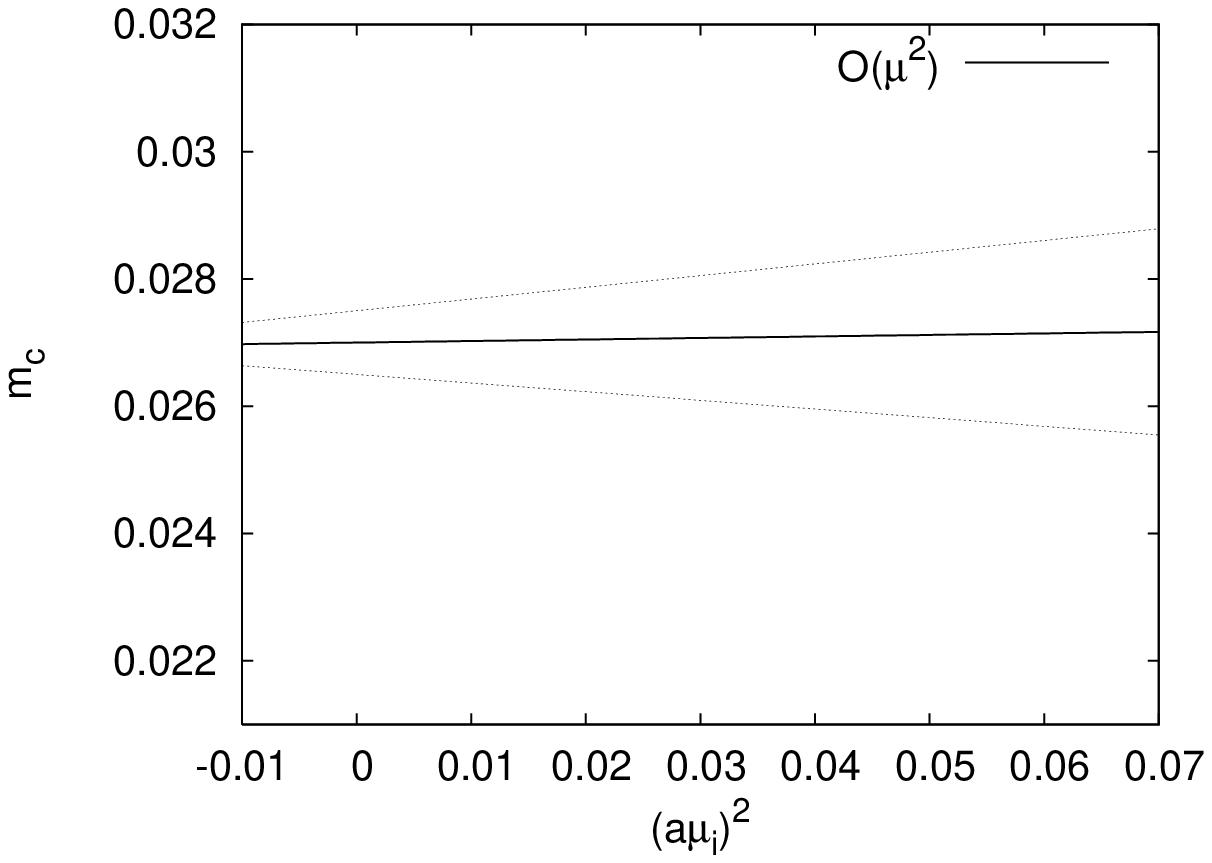}
\caption[]{\em 
Left: $B_4(am,a\mu)$ for different imaginary chemical potentials. The lines correspond to the
simultaneous fit of all data according to \eq (\ref{b4f}) and Table \ref{b4fits}, line 3. 
Right:  One sigma error band from a linear fit (Table \ref{b4fits}, line 3)
of the critical quark mass as a function of imaginary $\mu$.  }
\label{mcfig}
\end{figure}

Our data for five different values of imaginary chemical potential are shown in \fig\ref{mcfig} (left).
Remarkably, there seems to be negligible influence of the chemical potential.
The results of various simultaneous fits of all four curves are displayed in Table \ref{b4fits}. All fits are good, and none of the next-to-leading terms is significantly constrained. This is corroborated by
discarding all next-to-leading terms, which leads to a perfectly acceptable fit with $c_1'$ consistent with zero, as in the last line of Table \ref{b4fits}. 
\fig\ref{mcfig} (right) displays the error band coming from a linear fit
(Table \ref{b4fits}, line 3). 
Clearly, the slope $c_1'$ is very nearly zero.

\begin{table}[t]
\begin{center}
\hspace*{-1cm}
\begin{tabular}{|*{7}{r@{.}l|}l|}
\hline
\multicolumn{2}{|c|}{$m^c_0$} &
\multicolumn{2}{c|}{$b_{10}$} &
\multicolumn{2}{c|}{$b_{20}$} &
\multicolumn{2}{c|}{$c_1'$} &
\multicolumn{2}{c|}{$c_2'$} &
\multicolumn{2}{c|}{$C$} &
\multicolumn{2}{c|}{$\chi^2/{\rm dof}$} \\
\hline
0&0262(7) & 13&3(1.4) & -91&6(143.5) & -0&079(47) & -1&6(1.0) & -2&1(3.5) & 0&90\\
0&0263(6) & 13&9(0.6)  & \none    & -0&075(42)  & -1&35(0.73)  & \none & 0&82 \\
0&0270(5) & 13&6(0.6) & \none     & -0&0024(160) & \none & \none & 0&93 \\
0&0271(3) & 13&6(0.6) & \none     & \none               & \none  &\none  & 0&88\\
\hline
\end{tabular}
\caption{ \label{b4fits} 
  {\em 
Fitting $B_4(am,a\mu)$ to a Taylor expansion to different orders in the independent variables, according to \eq (\ref{b4f}). The numbers of d.o.f.~are 14,16,17,18, respectively.}}
\end{center}
\end{table}
The final result is then obtained by employing \eq(\ref{conv}) to convert to continuum units.
The second factor in \eq(\ref{conv}) is 1.077(2), close to 1.
In the first factor, the term $B$, which describes the variation of $T_0(\mu)$
with real $\mu$ and is thus negative, 
reinforces the negative trend of $c'_1$, to yield
\be
\frac{m^c(\mu)}{m^c(\mu=0)}=1 - 0.7(4) \left(\frac{\mu}{\pi T}\right)^2+\ldots \quad .
\label{c0}
\ee
Hence, we arrive at the surprising result that the first order region in the phase diagram \fig\ref{schem} 
{\it shrinks} when a real chemical potential is switched on. This is contrary to the expected qualitative behaviour.

The reader will notice the large error on the coefficient in \eq (\ref{c0}). It is a conservative estimate and stems entirely from the larger error on $c_1'$. 
If one were to include a $\mu^4$-term, the previous conclusion would only be strengthened: the leading term gets more negative and a negative quartic term comes on top of it, cf.~Table \ref{b4fits}.

However surprising, our findings agree with preliminary results for the same lattice theory at finite isospin chemical potential,
which indicate that there too, the transition becomes weaker as the chemical potential is turned on~\cite{DKS}.
Finally, let us note that the same qualitative behaviour applies to the first order region in 
the heavy quark limit,
which has recently been shown to also shrink with real chemical potential \cite{skim}.

\section{The chiral critical surface for $N_f=2+1$ \label{sec:line}}

Having removed finite step size errors from the $N_f=3$ calculations, we 
proceed to map out the chiral critical line for non-degenerate
quark masses. 
All our simulations have been performed with the RHMC algorithm. Since $8^3\times 4$ lattices
proved to be large enough for our observable in the case of $N_f=3$, we use that lattice size to 
trace out the critical line,
performing another check of  finite volume effects at $a m_{u,d} = 0.015$ on a 
$12^3\times 4$ lattice.  

With two different quark masses in the theory, a technical question concerning 
the Binder cumulant arises. Obviously, $B_4$ can be constructed from the chiral condensate of either mass flavour. Universality guarantees that, in the infinite volume limit, either choice tends to the same universal value. However, in a finite volume there are corrections, and they are different for different operators. The corrections are minimised for that superposition of operators, which
corresponds most closely to the mapping of the QCD parameters onto the scaling fields of the effective 3d Ising model. It is well known that, even for the case of three degenerate flavours, this is
a superposition of the chiral condensate and the plaquette \cite{kls}, as well as higher dimension fermionic and gauge condensates. 

\begin{figure}
\begin{center}            
{\rotatebox{0}{\scalebox{0.7}{\includegraphics{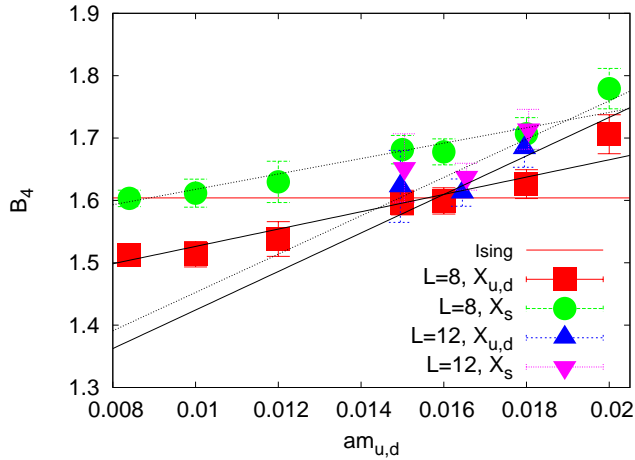}}}}
\end{center}
\caption[]{\em Finite size scaling of the fits to $B_4$ at fixed $am_s=0.075$, constructed from the light (solid line) and the heavy  (dotted line) flavour condensates for $L=8,12$. The light flavour observable has much smaller finite volume corrections.}
\label{pb1_pb2}
\end{figure}
Here, we do not attempt to construct an optimised observable by mixing in gauge condensates, but simply compare  the behaviour of $B_4$ constructed from condensates of different mass flavours, 
$X_i=\bar{\psi_i}\psi_i$, as shown in \fig\ref{pb1_pb2}. The observables $B_4(X_s),B_4(X_{u,d})$ constructed from the condensates of the heavy and light flavours, respectively, are observed to intersect the critical value at significantly different values of $am_{u,d}$. Nevertheless, comparison of results 
obtained at $L=8,12$  shows that this difference is rapidly disappearing on larger volumes. Moreover,
the common intersection point to which the results converge appears to be close to that obtained from $B_4(X_{u,d})$ on $L=8$,  indicating that the latter has far smaller finite volume corrections. 
This is not too surprising, as one would expect the scaling field corresponding to chiral
symmetry to be dominated by the lightest quark flavour.
Hence, in the following we will always work with $B_4(X_{u,d})$ constructed from light quark condensates.

\subsection{The critical line for $\mu_B=0$}
 
\begin{figure}      
\hspace*{-0.75cm}     
{\rotatebox{0}{\scalebox{0.65}{\includegraphics{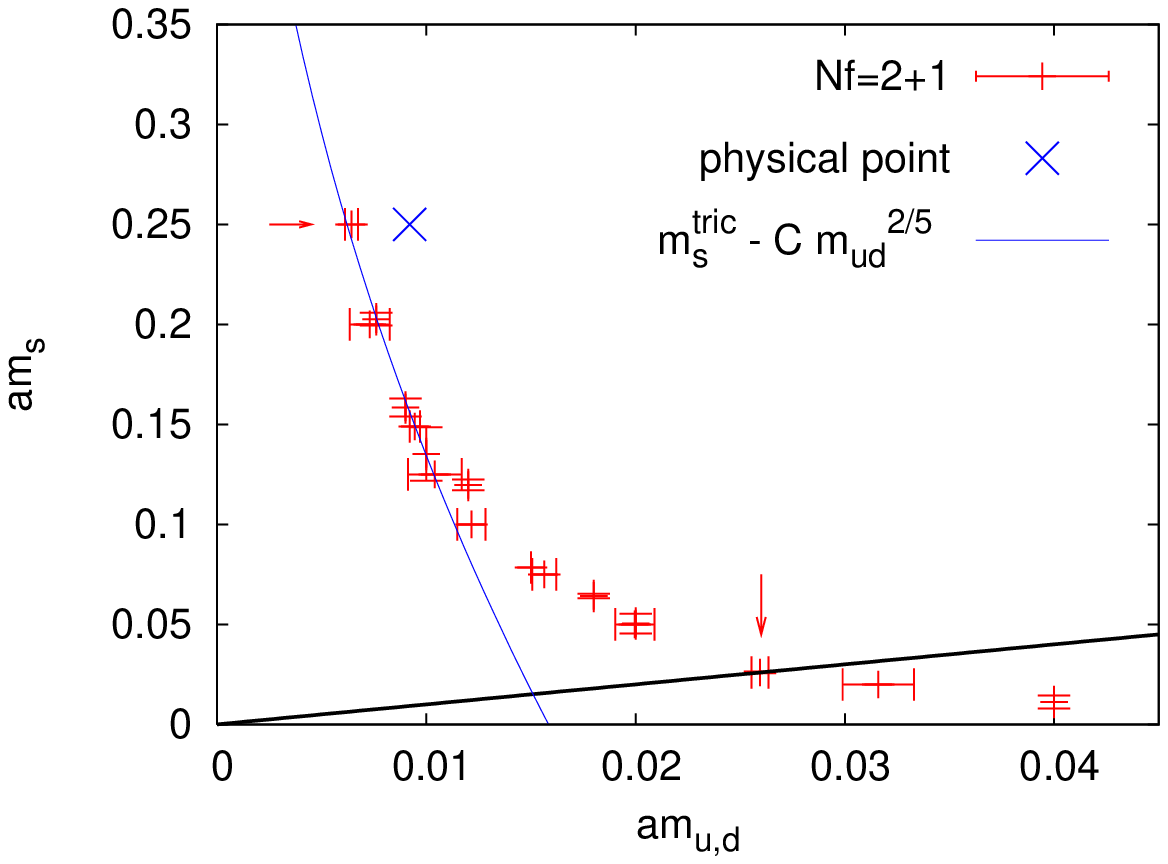}}}}
{\rotatebox{0}{\scalebox{0.65}{\includegraphics{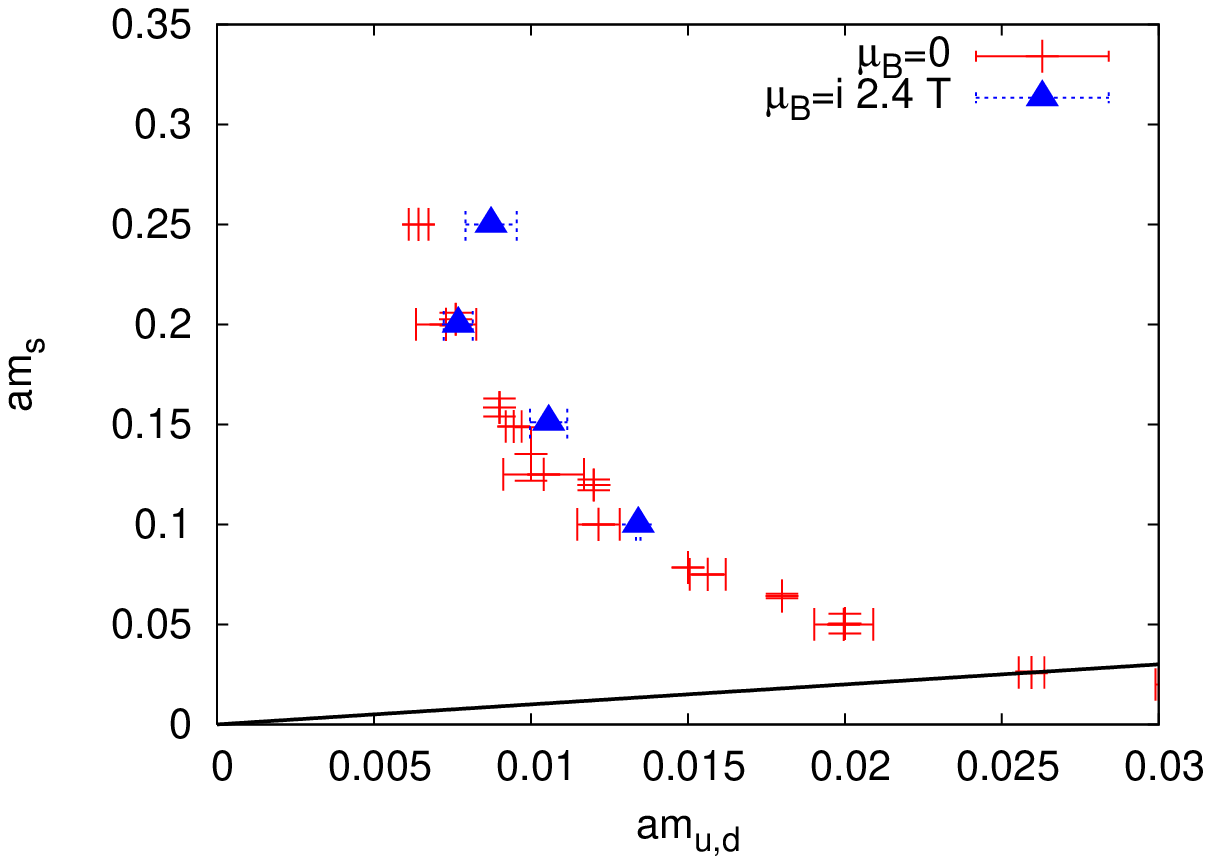}}}}
\caption[]{\em Left: The chiral critical line in the bare quark mass plane at $\mu_B=0$. The heavy line indicates the $N_f=3$ diagonal. Also shown is the physical point according to \cite{fk2}, and a fit
to extrapolate the line to a possible tricritical point on the $m_s$-axis. The arrows mark the points
where $T=0$ simulations were performed to set the scale, Sec.~\ref{sec:teq0}. 
Right: Comparison of the critical line at $\mu_B=0$ and $\mu_B/(iT)=2.4$.}
\label{m1m2c}
\end{figure}

By fixing $m_s$ and scanning in $m_{u,d}$ (at least 4 values), the critical light quark mass for that choice of $m_s$ is determined by interpolation, analogously to the three-flavour case. This is repeated for other values of $m_s$, resulting in the sequence of critical points $m_{u,d}^c(m_s)$ displayed in \fig\ref{m1m2c}, left. As in the three-flavour case, every
critical point appearing in this figure consists of at least 800k RHMC trajectories.

There are several features of \fig\ref{m1m2c} worth discussing. An interesting observation 
concerns the behaviour of the function $m_s^c(m_{u,d})$ in the neighbourhood of the three flavour critical point. If $B_4$ is constructed from gauge condensates and neglecting the change in the 
pseudo-critical coupling $\beta_0(m_{u,d},m_s)$ with the quark masses, 
a Taylor expansion around the symmetric critical mass $m^c_0$ yields for the line of constant (critical) $B_4$ the leading order result \cite{kls} 
\be
m_s=m^c_0-2(m_{u,d}-m^c_0),
\label{lo}
\ee 
i.e.~the critical line should pass through the symmetrical point with slope -2. 
In contrast, our data extracted from $B_4(X_{u,d})$ exhibit a different slope,
see \fig\ref{m1m2c} (left).
This underlines again the importance of choosing an appropriate observable for finite volume computations. A Taylor expansion of $B_4$ as in \eq (\ref{lo}) is only possible on finite volume, 
but expanding $B_4(X_{u,d})$ would yield additional non-perturbative contributions to \eq (\ref{lo}).
We thus conclude that \eq (\ref{lo}) does {\it not} describe the critical line, not even in the immediate
neighbourhood of $N_f=3$.

Another interesting question is how the critical line continues to even smaller light quark masses.
If the chiral limit of the $N_f=2$ theory exhibits O(4) universality, then the critical line hits the axis $m_{u,d}=0$ in a tricritical point at some finite strange quark mass value $m_s^{tric}$ \cite{ptgen}.
Whether this scenario is realized or not is an issue not yet settled (cf.~the discussion and references
in \cite{op}). Among the most recent publications using staggered fermions, one favors a first order scenario for the chiral limit \cite{pisa} while the other supports the O(4) scenario \cite{ks1}.
With our current data, we are unable to decide this question, but 
we can check for consistency with the O(4) scenario, 
which implies mean-field exponents  near the tricritical point $(m_{u,d}=0,m_s=m_s^{tric})$.
Indeed our data support a fit to the $\sim m_{u,d}^{2/5}$ approach to the chiral limit, as shown in 
\fig \ref{m1m2c}, left, predicting the tricritical point to be at $am^{tric}_s \sim 0.7$ or $m_s^{tric}/T\sim 2.8$.
Note however that $(i)$ our $N_t=4$ lattice is very coarse ($a \sim 0.3$ fm), and 
$(ii)$ our spatial volume becomes rather small as $m_{u,d}$ is reduced:
for the uppermost point in \fig \ref{m1m2c}, left, 
corresponding to the physical strange quark mass, $m_\pi L \sim 1.7$ only.
Thus, our systematic error might be rather large. Nevertheless, we have strong
indications that $m_s^{tric}$ is significantly larger than the physical strange quark mass.

\subsection{The chiral critical line and the physical point \label{sec:teq0}}

The most important question regarding the critical line is, of course, its location relative to the 
physical point of QCD. So far, all known lattice data are ``consistent'' with the physical point being in the crossover region. This is also the result found by a simulation at physical quark masses
in \cite{fk2}. However, these results were obtained by the R-algorithm, and we have seen in the three flavour case that significant shifts in the critical quark masses can arise due to step size errors. 

In order to estimate the location of our critical line in physical units, we have therefore
performed zero temperature simulations with bare quark masses corresponding to two points on the critical line. One corresponds to the three flavour theory and the other to the point with roughly physical strange quark mass, as indicated by the arrows in \fig\ref{m1m2c} (left). 
The parameters of the simulations are given in Table \ref{teq0}, together with
the measured meson masses. In both cases, the lattice
size was $12^3\times 24$, and about 400 configurations were analyzed.

\begin{table}
\begin{tabular}{|cc|ccc|cc|}
\hline
$(am_{u,d}, am_s)$ & $\beta$ & $a m_\pi$ & $a m_K$ & $a m_\rho$ & $m_\pi/m_\rho$ & $m_K/m_\rho$ \\
\hline
(0.0265,0.0265) & 5.1374 & 0.420(1) & 0.420(1) & 1.383(7) & 0.304(2) & 0.304(2)\\
(0.005,0.25) & 5.1857 &  0.2109(1) & 0.8915(1) & 1.398(16) & 0.151(2) & 0.638(8) \\
\hline
\end{tabular}
\caption[]{\label{teq0} \em
Parameters and meson masses for the low-temperature ($12^3\times 24$)
simulations performed to set the scale. The mass ratios of the second line imply that this point 
corresponds to $m_s=(m_s)_{phys}, m_{u,d}<(m_{u,d})_{phys}$.}
\end{table}

Setting a physical scale along the critical line is a tricky problem.
Neither of our simulation points matches the physical $(m_{u,d},m_s)$ point,
so that strictly speaking one cannot match to any real world observable.
Doing so anyway,  different observables inevitably give different values for the lattice
spacing.
A measurement of the $q\bar{q}$ force via elongated Wilson
loops gives $r_0/a = 1.85(2)$ and 1.87(2) respectively, where $r_0=0.5$ fm is the Sommer
scale. This amounts to $a(r_0)\approx 0.27$ fm in both cases.
On the other hand, matching the $\rho$-mass to its physical value gives a lattice spacing which is
by 20\% larger. 
Note, however, that a difference of similar magnitude has been observed in \cite{fk2} on the physical point. This suggests that the greater part of this difference is due to cut-off effects rather 
than to the deviation from physical parameters. 

It thus appears safer to avoid setting an absolute scale altogether, and instead compare the meson mass ratios from Table \ref{teq0} with their values at the physical point,
$(m_\pi/m_\rho)_{phys}=0.18$ and $(m_K/m_\rho)_{phys}=0.645$.  We thus conclude that our $N_f=2+1$ point
on the critical line indeed corresponds to the physical Kaon mass and to pions lighter than physical.
In other words, the physical point is on the crossover side of the critical line. 

\begin{figure}[t]
\vspace*{-0.5cm}
\begin{center}
\includegraphics*[width=0.6\textwidth]{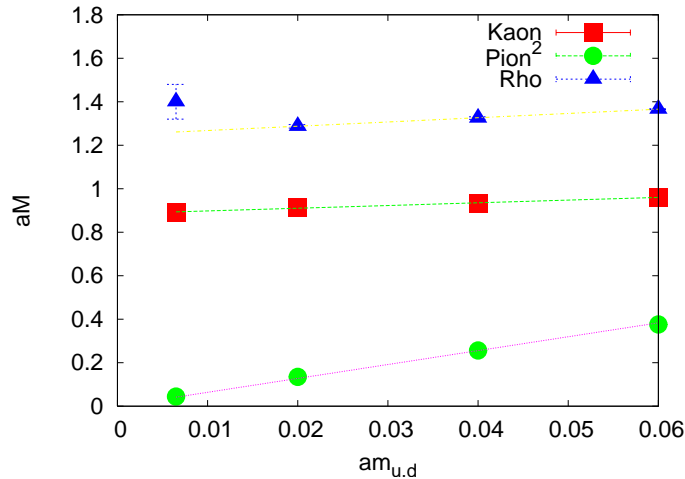}
\caption[]{\em
Comparison of our $N_f=2+1$ meson masses $aM$ (leftmost points) with those
obtained in Ref.\cite{fk2} with the R-algorithm. Good consistency is apparent.}
\label{masses_compare}
\end{center}
\end{figure}

It is interesting to compare our $N_f=2+1$ meson masses, in 
lattice units, with those of \cite{fk2}, which used the same strange quark
mass, and the R-algorithm at almost the same inverse coupling $\beta=5.19$.
The comparison is shown in \fig \ref{masses_compare}, where the
straight lines are fits to the data of \cite{fk2} only. Good consistency
is apparent, showing that the R-algorithm step size error for the meson
masses is small, for the parameters considered in \cite{fk2}
\footnote{
Note also, that there are preliminary results by Z.~Fodor and S.~Katz
(https://www.bnl.gov/sewm/) using finer lattices and the exact RHMC algorithm,
which confirm that the physical quark masses give a finite-temperature
crossover.}. The figure also shows $m_K/m_\rho$ to be practically independent of 
$m_{u,d}$, thus affirming our conclusion above.

Finally, the fact that the lattice spacing varies little between our two simulation
points, implies that $T_c$ itself does not change much as one moves
along the critical $(m_{u,d},m_s)$ line. This is in agreement with
model calculations \cite{effrev}.

\subsection{The critical line for $\mu_B/(iT) = 2.4$ and the critical surface}

We have also run a second set of simulations with an imaginary chemical potential $\mu_B/(iT) = 2.4$, in order to determine how the critical line shifts with baryon density. These data are shown in 
\fig\ref{m1m2c} (right), in lattice units. We observe that the shift in the critical line is very small, despite the sizeable value of the chemical potential. Within two sigma the line 
is consistent with its $\mu_B=0$ counterpart.
Moreover, to the extent that there is a displacement,  it shows a trend to lie to the right of the zero density line. 
This qualitative observation is in accord with 
our earlier finding in the three flavour case (\eq \ref{c03lat}) 
that $c'_1 \sim 0$ or slightly negative: in lattice units, the first order
region tends to expand slightly as an imaginary chemical potential is turned on
(see \fig \ref{mcfig} (right)).

Similarly to the $N_f=3$ case, one expects the data along the whole line to be described
by the leading term in the Taylor expansion,
\be
am_{u,d}^c(am_s,a\mu) = am_{u,d}^c(am_s,a\mu=0) +  c'_1(am_s) \left( a\mu \right)^2+\ldots
\ee
where now $c'_1$ depends on $am_s$.
Conversion to continuum units proceeds as for $N_f=3$, by determining the equivalent of 
\eqs (\ref{nf3temp},\ref{conv}) for $N_f=2+1$. 
We do this for fixed physical strange quark mass $am_s=0.25$ and scanning in $m_{u,d}$,
which now plays the role of the variable quark mass, and find $am^c_{u,d}\sim 0.0064$.
For the variation of the pseudo-critical coupling, \eq (\ref{beta}), we obtain the coefficients given in Table
\ref{fits2}, leading to $A=1.90(13)$ and $B=-0.49(1)$ in this case.
\begin{table}[t]
\begin{center}
\hspace*{-1cm}
\begin{tabular}{|*{5}{r@{.}l|}l|}
\hline
\multicolumn{2}{|c|}{$c_{00}=\beta_0(0,m^c_{u,d})$} &
\multicolumn{2}{c|}{$c_{10}, (\mu^2)$} &
\multicolumn{2}{c|}{$c_{20}, (\mu^4)$} &
\multicolumn{2}{c|}{$c_{01}, (m)$} &
\multicolumn{2}{c|}{$\chi^2/{\rm dof}$} \\
\hline
 5&1838(3) & 0&572(9)  & \none    & 1&75(13)   & 1&5 \\
\hline
\end{tabular}
\caption{ \label{fits2} 
  {\em 
Fit of the Taylor expansion $\beta_0(am_{u,d},a\mu)$,
\eq (\ref{beta}), to our data for fixed  $am_s=0.25$ and $am^c_{u,d}\sim 0.0064$. }}
\end{center}
\end{table}
These are similar in magnitude to the three-flavour
values. Note that $c'_1\approx 0$ means that $m_{u,d}^c/T_c$ remains constant as
the chemical potential is turned on.
The decrease of the pseudo-critical temperature with real $\mu$, given by $B$,
is then the dominant effect. It dictates that $m_{u,d}^c$ {\em decreases}
when  a real chemical potential is turned on.
In other words, the first order region {\em shrinks}.

While our data for $N_f=2+1$ and small quark masses have larger errors which do not yet allow to constrain $c_1(m_{u,d})$ quantitatively, 
\fig\ref{m1m2c} and \eq(\ref{conv}) leave little doubt that this coefficient is going to be negative along the whole upper part of the line. 
We thus arrive at the conclusion that, for the lattice spacing considered here, the curvature of the critical surface at $\mu_B=0$ is negative, and the first 
order region is shrinking when a real chemical potential switched on.

\begin{figure}[t]
\vspace*{-0.5cm}
\hspace*{-1.5cm}
\includegraphics*[width=0.6\textwidth]{3dphasediag_4.eps}\hspace*{0.25cm}
\includegraphics*[width=0.5\textwidth]{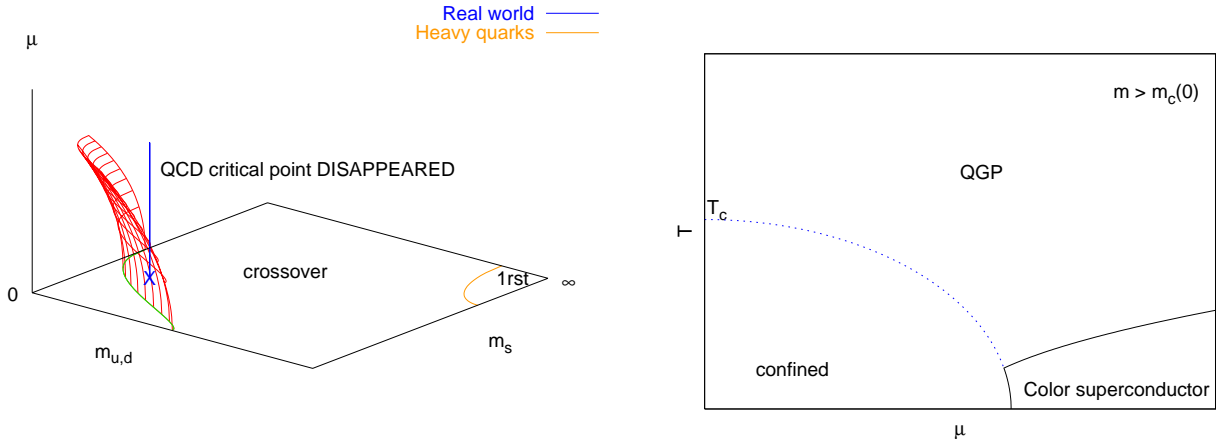}
\caption[]{\em
For $dm_c(\mu)/d\mu^2 <0$, there is no critical point at all, the dotted line on the right is merely a crossover. 
Any additional critical structure would not be continuously connected to that at $\mu=0$. }
\label{nons}
\end{figure}

\subsection{An alternative scenario for the QCD phase diagram}

Let us take our results at face value for a moment and consider the implications if such a qualitative result also holds in the continuum limit. 
This leads to a scenario for the $(T,\mu)$ phase diagram which is at odds with common expectations.
We find that the first order region in a plane of constant $\mu_B$ is actually shrinking with growing real 
$\mu_B$. If the physical point is in the crossover region at $\mu_B=0$, then switching on a chemical potential will not lead to an intersection with the critical surface 
as long as the latter is well described by its curvature at $\mu_B=0$,
i.e. for $\mu/T \lsim 1$, or equivalently $\mu_B\lsim 500$ MeV. 
In the absence of any additional (and so far unknown) critical structure,
there would then be no critical point or first order phase transition at all.
The $(T,\mu)$ phase diagram of physical QCD would then only have the possible transition line separating the superconducting phase from nuclear matter, as in \fig \ref{nons} (right).

\begin{figure}[t]
\vspace*{-0.5cm}
\begin{center}
\includegraphics*[width=0.45\textwidth]{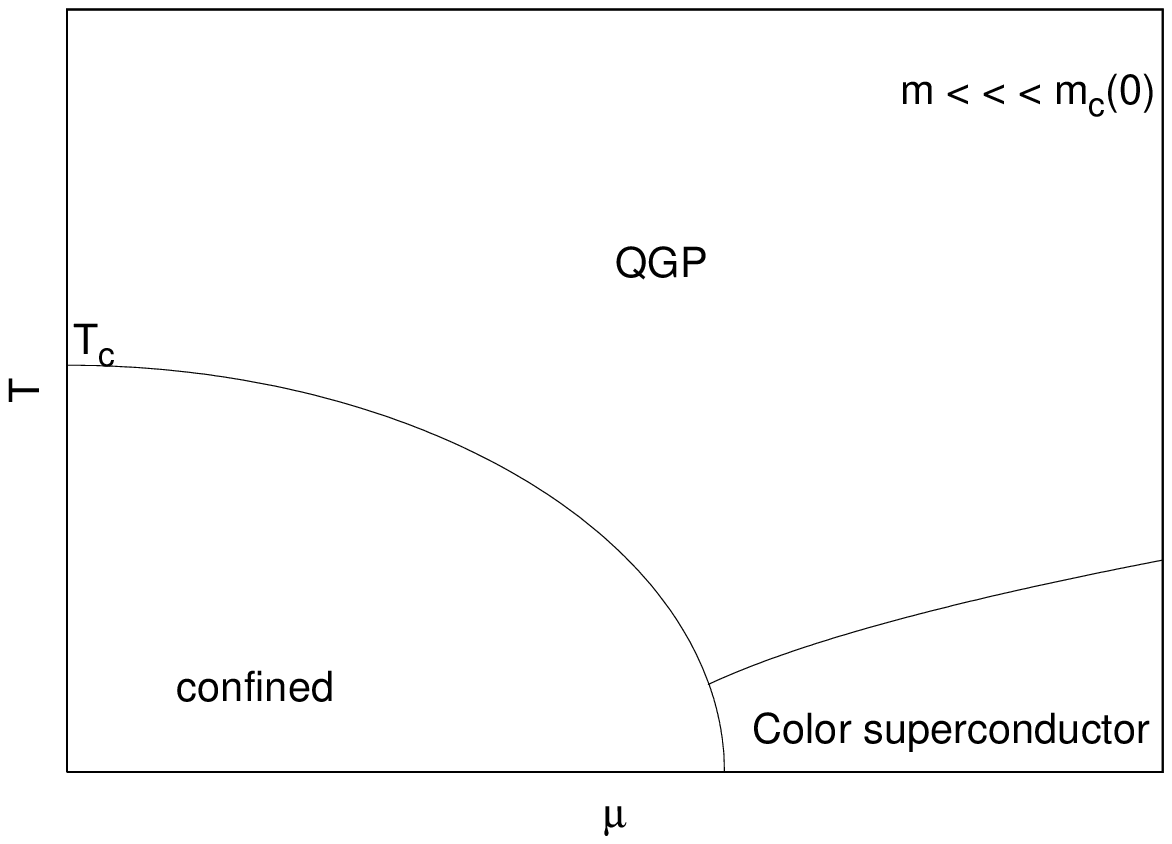}
\includegraphics*[width=0.45\textwidth]{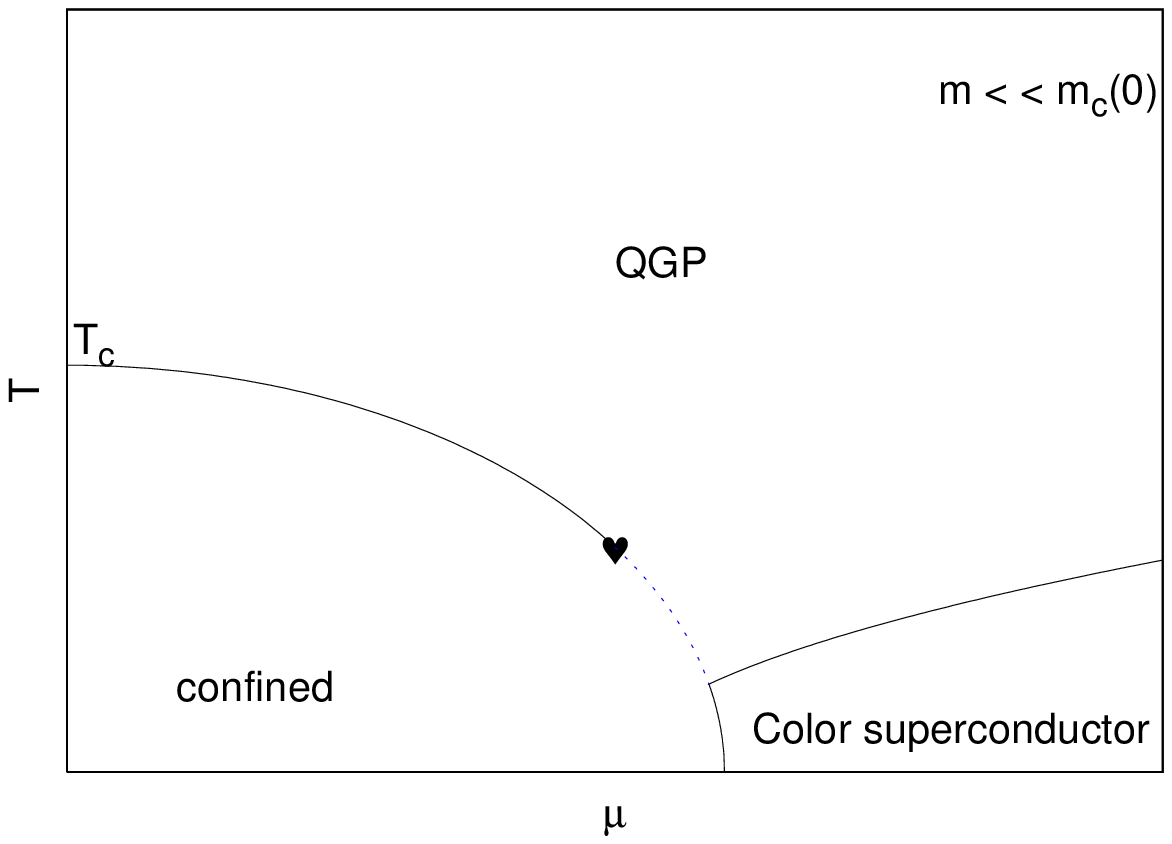}
\includegraphics*[width=0.45\textwidth]{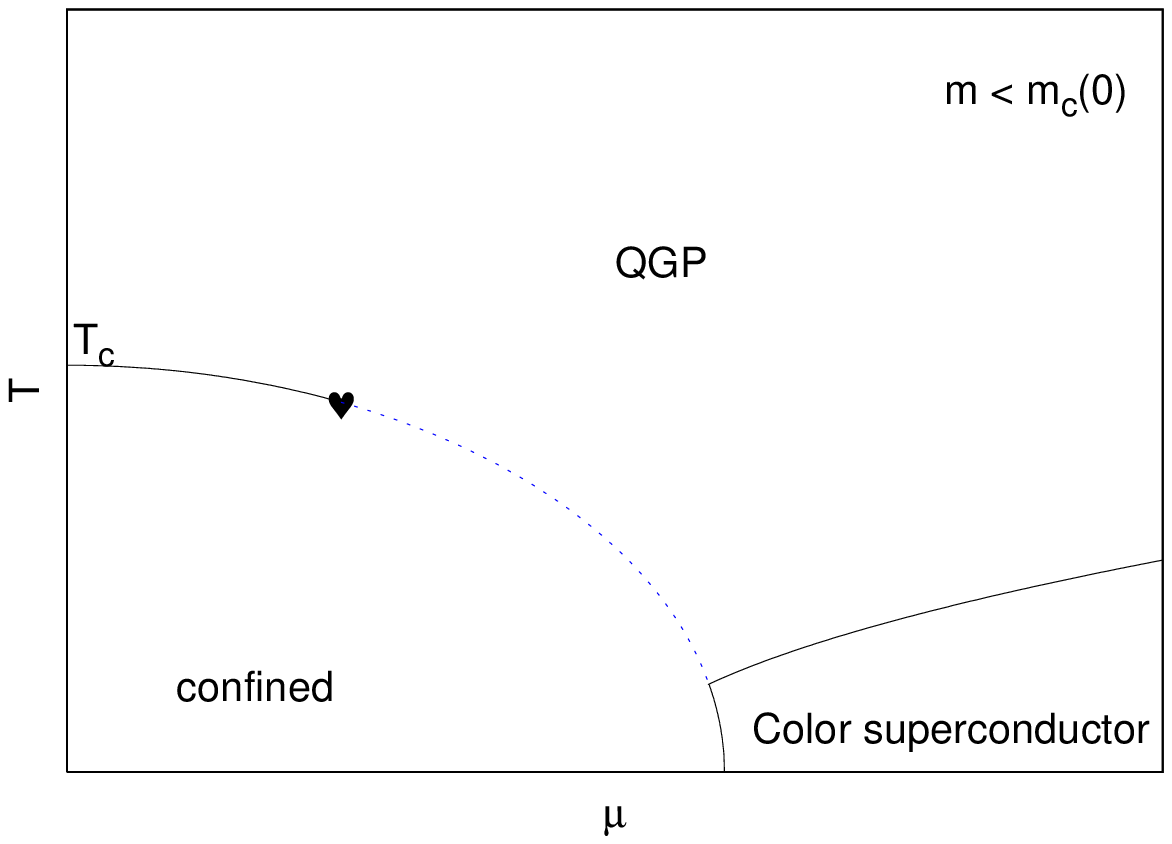}
\includegraphics*[width=0.45\textwidth]{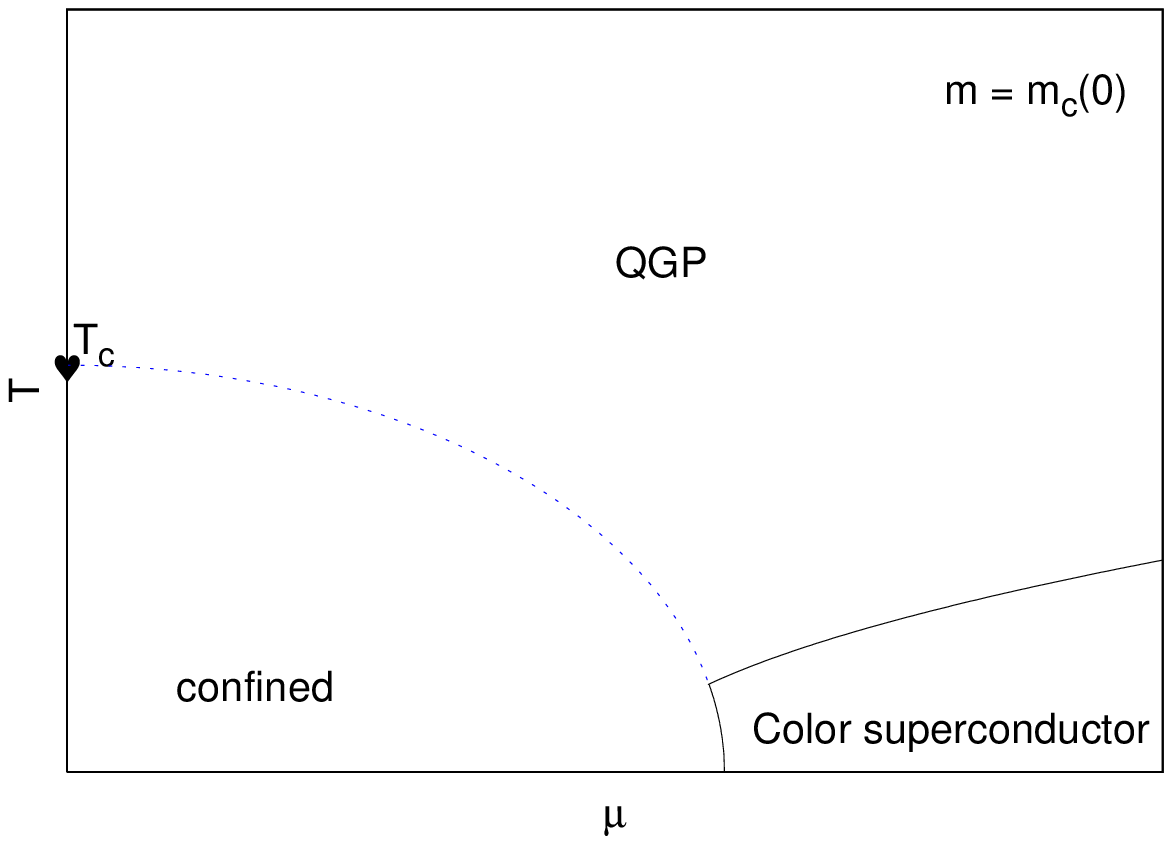}
\caption[]{\em The $N_f=3$ phase diagram as a function of quark mass. The critical endpoint now is moving
to smaller $\mu^c$ with growing quark mass, due to the negative sign of $c_1$ in Eqs.~(\ref{c0},\ref{c03}).}
\label{scenario}
\end{center}
\end{figure}

Note that this scenario is perfectly consistent with all universality arguments and the known results 
for $\mu=0$. This can be illustrated in the three flavour theory by
considering the change of the $(T,\mu)$-diagram with quark mass, as depicted in \fig\ref{scenario}. 
All boundary conditions are met, in particular there is a first order phase transition at $\mu=0$ 
for quark masses smaller than $m^c_0$.
However, according to the negative sign for $c_1$ in Eqs.~(\ref{c0},\ref{c03}), the critical
endpoint is now moving to the left with growing quark masses, until it disappears entirely. 

It is natural to ask how reliable this unexpected scenario is regarding systematic errors.
We have seen in Sec.~\ref{sec:teq0} that $N_t=4$ lattices are very coarse, and we have discussed
the enormous sensitivity of the critical values of the mass parameters $(m_{u,d},m_s)$ to cut-off effects
\cite{p4_vs_KS}. It is therefore expected that the values for $m_c(\mu_i)$ shift once this analysis is repeated on finer lattices and/or with improved actions. The crucial question is whether the continuum
extrapolated slope $c'_1$ in \fig \ref{mcfig}(right) will be positive or negative 
and how it will balance out against the
contribution from the pseudo-critical temperature. Or, equivalently, how the ordering of
the zero and finite $\mu$ critical lines in \fig\ref{m1m2c} turns out in the continuum limit once physical units are used.
In particular, our continuum conversion is sensitive to the non-perturbative beta function.
A look at \eq (\ref{conv}) gives a quick estimate of what is needed for a positive $c_1$.
Either $c_1'$ would have to be positive of the order $\sim am^c_0 N_t^2/\pi^2$, or $A$ has to grow
by a factor larger than 10 on the way to the continuum.    
Thus, while we presently cannot rule out that the picture reverts back to the standard scenario in the continuum limit, the opposite is obviously also possible as suggested by our data.

Finally, all of our arguments are based on the simplest scenario, in which the finite $\mu$ critical point
of physical QCD is continuously connected to a critical point for some other mass values at $\mu=0$.
We cannot exclude a more complicated possibility, where the phase boundary of
the color superconducting phase (see \fig\ref{scenario}) would distort as a 
function of the quark masses, and give birth to a critical point distinct from 
the one we study, and which would survive for physical quark masses.
This would correspond to a scenario with an additional critical surface in Fig.\ref{nons} (left) 
above the one we studied here.
We thus conclude that even the qualitative features of the QCD phase diagram cannot be regarded as settled yet.

\subsection{The finely tuned critical end point}

While based on our present data we are unable to make a reliable 
quantitative prediction for the location of the critical point for physical QCD, we have obtained important qualitative information regarding its quark mass dependence. Irrespective of the continuum extrapolated
sign, all our evidence is that the absolute value of $c_1$ in \eq(\ref{c1}) will be $\sim O(1)$ as naturally expected, while the effect of  subleading terms is small up to $\mu_B\approx 500$ MeV. This means that the critical quark
masses are very weakly varying functions of the chemical potential, which is in line with the 
corresponding behaviour of the pseudo-critical temperature or, indeed, the equation of state.
Consequently the inverse function $\mu_c(m_{u,d})$ is very strongly varying with quark mass, i.e.~the
finite $\mu$ critical surface emerges very steeply from the $\mu=0$ critical line in \fig\ref{schem}.
Hence, even if in the continuum limit the conventional scenario with positive curvature is realized,
the precise location of the critical end point will be exceedingly quark mass sensitive. A simple estimate
using $c_1=1$ shows that, in order to have $\mu_B^c\lsim 400$ MeV, the physical point has to lie within $\lsim 5\%$ of the chiral critical line, i.e.~the physical quark masses would be fine tuned.
While there is nothing forbidding such a situation,
it appears rather unnatural. Moreover, 
if realized in nature, it would make a quantitative determination of $\mu^c_B$ through simulations exceedingly difficult.
(For example, one might even need to treat the $u$- and $d$-quarks as non-degenerate).

\section{Discussion and conclusions \label{sec:sys}}

As we have demonstrated in the preceding sections, step size errors are eliminated from our calculations. Finite volume effects are under control in the $N_f=3$ theory
and for $am_{u,d}\geq 0.015$ in $N_f=2+1$, where we have performed our finite volume check. 
Note that this corresponds to pion masses larger than physical. For the point on the critical line with the lightest quark mass considered, we only have $m_\pi L \sim 1.7$, and finite volume effects are to be expected. Ideally the part of the critical line in the neighbourhood of the physical point should be checked on a larger lattice.

However, by far the largest source of uncertainty is due to the coarse lattice spacing $a\sim 0.3$ fm,
as evidenced by several aspects of these computations.
Strong cut-off effects reveal themselves when attempting to set a physical scale for the problem.
Moreover, 
a change in the discretization of the Dirac operator on a lattice this coarse 
can change the pion mass corresponding to the second-order
transition by a factor $\sim 4$~\cite{kls,p4_vs_KS} at $\mu_B=0$.
Finally, it has recently been pointed out that staggered simulations at finite $\mu_B$ suffer from additional discretization errors compared to $\mu_B=0$ \cite{stagmu} when $N_f \neq 4$, due to the eigenvalue structure when taking the fourth root of the determinant. 
For simulations at imaginary $\mu_B$, the eigenvalues are pure imaginary, and this additional error
is of $O(a^2)$, with possibly a large coefficient. A safe strategy thus is to first take the continuum limit of imaginary $\mu_B$ simulations, and then continue to real $\mu_B$. For reweighting approaches at real $\mu_B$ one even expects $O(a)$ errors.
 
In interpreting our findings and comparing with other work, it is important to take systematic uncertainties into account.
Given the cut-off effects, the sensitivity of the critical point to step size errors and, most notably, to the quark mass, it is clear that the discrepancy between our findings and those of \cite{fk2} is nothing remarkably unusual, but merely reflects the large and different systematic uncertainties afflicting
these calculations. 
In particular, in \cite{fk2} the quark masses $am_q$ were held fixed in lattice units while $a\mu$ was
increased. 
Equivalently, $m_q/T_0$ was kept fixed. However, $T_0$ decreases under the
influence of a chemical potential, in a manner similar to \eq \ref{nf3temp},
so that the quark masses at the critical endpoint in \cite{fk2} are about 5-10\% smaller than physical. 
This small deviation from a line of constant physics has a large impact on the
location of the critical point, because of the high sensitivity of the latter on
quark masses\footnote{In our approach, the conversion from lattice units
\eq (\ref{c03lat}) to physical units \eq (\ref{c03}) changes the coefficient $c'_1$,
which is nearly zero, to $c_1$, which is negative.}.
The effect is to artificially move the critical point to smaller
chemical potentials. The shifted masses may even reside in the first order
region, causing a critical point to be found even if in fact there is none,
consistently with the scenario discussed here.
 
Our study of the chiral critical surface also suggests that one cannot draw conclusions for physical QCD from
simulations of the critical point in $N_f=2$ QCD \cite{gg,swabi}.
This  becomes clear when considering \fig \ref{schem1},
which describes the $\mu=0$ expectations.
If one moves from the physical point upwards by increasing $m_s$ to infinity,
the distance to the critical line increases considerably.
Given the high sensitivity of the critical chemical potential $\mu^c$ to this
distance (i.e. the small curvature of the chiral critical surface in
\fig \ref{schem}) which we observe, one should expect large differences
for $\mu^c$ between the $N_f=2+1$ and $N_f=2$ theories
\footnote{Moreover, for $N_f=2$ one expects a tricritical point 
$\mu^{tri}(m_{ud}=0)$, with mean-field critical exponents which govern
the analytic form of the critical line $\mu(m_{ud},m_s=\infty)$, in contrast
to the $N_f=2+1$ case.}.

Resolving these various systematic issues and deciding which scenario for the $(T-\mu)$
phase diagram is realized in nature thus urgently requires further investigations of the $N_f=2+1$ theory with exact algorithms on finer lattices. 
Among the various finite $\mu$ approaches, our imaginary $\mu$ simulations require comparatively moderate computer resources to achieve this goal.

\vspace*{1cm}
\noindent
{\bf \large Acknowledgements:} We are grateful to K.~Rummukainen, Y.~Shamir and B.~Sve-titsky for discussions. We thank M.~Stephanov for a thorough reading and many discussions. We also thank the Minnesota Supercomputer Institute, the RCNP Osaka and the HLRS Stuttgart for computer resources.

\end{document}